# Smart Bengali Cell Phone Keypad Layout

Rezwana Sharmeen, Md. Abul Kalam Azad, Shabbir Ahmad and S. M. Kamruzzaman[†]
Department of Computer Science & Engineering,
International Islamic University Chittagong, Chittagong, Bangladesh.
[†] Department of Computer Science & Engineering, Manarat International University, Dhaka, Bangladesh.
{ r_sharmin_79, azadarif, bappi_51, smk_iiuc}@yahoo.com

**Abstract**

*Nowadays cell phone is the most common communicating used by mass people. SMS based communication is a cheap and popular communication method. It is human tendency to have the opportunity to write SMS in their mother language. Text input in mother language is more flexible when the alphabets of that language are printed on the keypad. Bangla mobile keypad based on phonetics has been proposed earlier. But the keypad is not scientific from frequency and flexibility point of view. Since it is not a feasible solution in this paper we have proposed an efficient Bengali keypad for cell phone and other cellular device. The proposed keypad is based on the frequency of the alphabets in Bengali language and also with the view of structure of human finger movements. We took the two points in count to provide a flexible and fast cell phone keypad.*

## I. INTRODUCTION

WAP, chatting and messaging are considered to be the driving force for the next generation of mobile devices. New compact and extended function oriented cell phones from leading manufacturers are providing full pocket communication functionality. These cell phones allow users to browse the Internet, send and receive email, chatting and SMS, and handle personal data and information.

SMS and chatting service of the cell phone is widely used in our country. People of Bangladesh send SMS in Bengali language using English alphabets. But still there is no compact Bengali cell phone keypad layout [1,4]. Bangla is our mother language and 21st February has been declared as International Mother Language Day, Bengali language has made their place in the world as a renowned language. Furthermore Bengali is the mother tongue of more than 200 Millions of people. That's why a scientific Bengali cell phone keypad was very much cherished. In this paper we are proposing a Bengali keypad layout based on the frequency of the Bengali alphabets to increase the typing speed. Thus it will present a new mobile keypad layout taking into consideration the importance of mobile functionality in current world and making it more flexible to the users.

## II. PROPOSED ARCHITECTURE

We have taken into consideration the concept of flexibility of thumb movement according to the outline of medical science and frequency of the Bengali alphabet.

## III. IDENTIFYING THE FREQUENTLY USED

In this paper we have proposed a new one-handed one fingered cell phone keypad layout. Our keypad layout is designed by arranging the most frequently used alphabets in the most flexible keys. Flexible keys are those, which are flexible to use to the users. A key is flexible when it is reachable by the finger (thumb) without much pressurizing the physical structure and the internal joints of the thumb. A physical structure of the thumb is shown in Figure. 1.

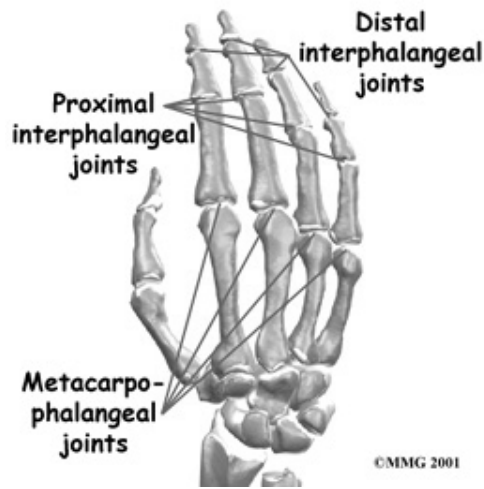

Figure 1: Physical structures of finger joints [7, 8, 9]

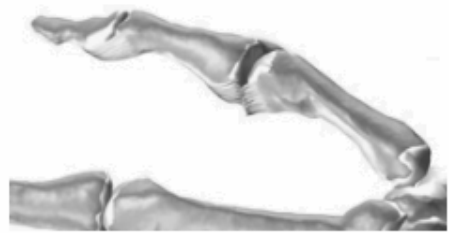

Figure 2: Physical structures of thumb joints [7, 8, 9]

There are mainly two joints in a thumb (1) interphalangeal (2) metacarpophalengeal Basically two types of thumb movements are required for pressing any key (1) Flexion (2) Extension. In case of Flexion movement the metacarpophalengeal joint is moved to forward direction whereas in case of extension metacarpophalengeal joint is moved to lateral direction, which is more painful and pressure creating. According to thumb joint movement principle, pressure on interphalangeal joint increases with the decrease in the joint angles. If the thumb movement pressurize the metacarpophalengeal joint to lateral direction then this movement pressurize the thumb movement and there by inconvenient and also a bit painful. The extension of the thumb is more painful and pressure creating than the decrease of angle in the interphalangeal joint.

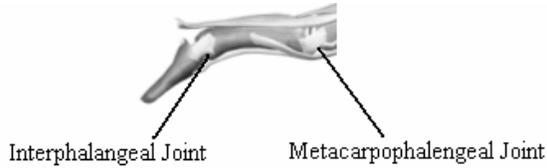

Figure 3: Physical structures of different joints [7, 8, 9]

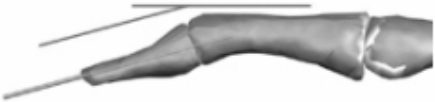

Figure 4: Interphalangeal joint movement [7, 8, 9]

Position of interphalangeal joint and metacarpophalengeal joint is shown in the following three figures: Figure 5, Figure 6 and Figure 7. In Figure 5 for pressing key 1 the interphalangeal joint is slightly bent and this angle of joint decreases as we proceed to press key 5 and key 9. While pressing key 9 there is extension of metacarpophalengeal to later direction.

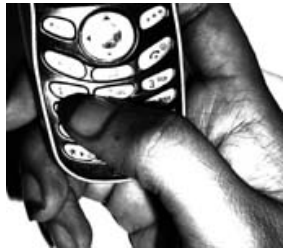

Figure 5: Pressing key 1

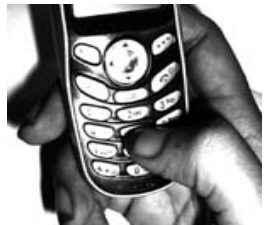

Figure 6: Pressing key 5

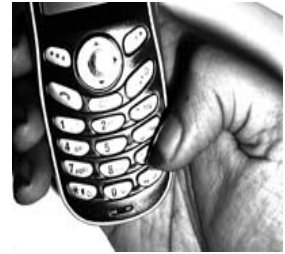

Figure 7: Pressing key 9

The pressure on the joints for thumb movements to press the cell phone keys are shown in table 1. Here IJ stands for Interphalangeal Joint and MJ stand for Metacarpophalengeal Joint.

Table 1: Analysis of thumb movement for each key press

| Key | IJ (Angle) | MJ (Direction) | Flexion | Extension |
|---|---|---|---|---|
| 1 | 120° | Forward | Yes | No |
| 2 | 110° | Forward | Yes | No |
| 3 | 80° | Lateral | No | Yes |
| 4 | 100° | Forward | Yes | No |
| 5 | 95° | Forward | Yes | No |
| 6 | 70° | Lateral | No | Yes |
| 7 | 80° | Forward | Yes | No |
| 8 | 70° | Lateral | No | Yes |
| 9 | 65° | Lateral | No | Yes |

From the above table the most flexible key in the first row is 1 then 2 as the angle decreases to 10 degree. Also from the above table the most flexible key in the second row is 4 then 5 as the angle decreases to 5 degree and the next flexible key in the third row is 7. The least flexible keys are consecutively 3,6,8,9 as these movements require extension to lateral direction and also the angles in the interphalangeal joint is less. The final arrangement of frequency is 1 > 2 > 4 > 5 > 7 > 3 > 6 > 8 > 9.

### IV. CATEGORIZING THE ALPHABETS ACCORDING TO FREQUENCY

To categorize the Bengali alphabets for SMS, Chatting purpose it was needed to develop a versatile database of Bengali words. For that reason, a database of nearly 9.5 Lac word combined to make the database. After that the frequency of the Bengali alphabet is calculated by the use of the conventional frequency calculation algorithm.

*Algorithm 1: Frequency (Alpha, AlphaLen, Data, DataLen)*

Here Data is a linear array of DataLen elements contaning Bengali alphabetic characters. Alpha is an array of positive integer of AlphaLen elements. This algorithm finds the frequency of the Bengali alphabet.

1. *Set Count =0*
2. *Repeat Step 3 and 4 while Count < AlphaLen*
3. *Set Alpha [ Count] = 0;*
4. *Set Count = Count + 1;*

*[End of Step 2 Loop]*
5. Repeat Step 6,7 and 8 while Count t< DataLen
6.     Set C = Data [ Count]
7.     Set Count = Count + 1;
8.     set Alpha [ C] = Alpha [C] + 1;
*[ End of Step 5 Loop]*
9. Exit

The Bengali alphabets are arranged in table 2 according to the decreasing order of the frequency. Here SL stands Serial Position and ALPH stands for Alphabet.

Table 2: Arrangement of Bengali alphabets according to frequency

| SL | ALPH | SL | ALPH | SL | ALPH |
|---|---|---|---|---|---|
| 1 | া | 21 | এ | 41 | থ |
| 2 | ে | 22 | য | 42 | জ |
| 3 | র | 23 | জ | 43 | ৈ |
| 4 | ি | 24 | থ | 44 | চ |
| 5 | ক | 25 | ূ | 45 | ‌ |
| 6 | ন | 26 | গ | 46 | ঘ |
| 7 | ব | 27 | ধ | 47 | ঔ |
| 8 | স | 28 | ণ | 48 | ঝ |
| 9 | ল | 29 | ছ | 49 | ড |
| 10 | ত | 30 | ট | 50 | ঞ |
| 11 | প | 31 | , | 51 | উ |
| 12 | য় | 32 | ভ | 52 | ী |
| 13 | দ | 33 | ঃ | 53 | ঈ |
| 14 | ম | 34 | ড় | 54 | ও |
| 15 | হ | 35 | ং | 55 | খা |
| 16 | শ | 36 | চ | 56 | ঢ |
| 17 | ী | 37 | ফ | 57 | ্ |
| 18 | অ | 38 | ও | 58 | ঘ |
| 19 | ্ | 39 | ই | 59 | ৃ |
| 20 | ্ | 40 | ঠ | 60 | ঊ |

From the calculation we have seen that the symbols ( া, ে, ি ) the shorted form of the vowels are more frequently used than the vowels. We have devoted two keys for the symbols and vowels. The Bengali consonants are arranged according to their frequency of occurrence in other keys. The most frequently used consonants are shown in table 3 according to the percentage of occurrence of each consonant in our sample inputs. Percentage refers to the percentage of occurrence of each Bengali consonant in our sample inputs. Here CON stands for Consonant and PER stands for Percentage.

Table 3: Frequency of the Bengali consonants

| CON | PER | CON | PER | CON | PER |
|---|---|---|---|---|---|
| i | 12.89% | k | 9.03% | D | 1.54% |
| K | 12.65% | L | 8.67% | V | 1.13% |
| b | 12.10% | R | 8.07% | L | 1.09% |
| e | 12.09% | _ | 7.85% | P | 1.03% |
| m | 12.03% | N | 7.64% | N | 0.97% |
| j | 12.01% | a | 6.98% | o | 0.65% |
| Z | 11.98% | Y | 6.02% | S | 0.43% |
| c | 11.66% | Q | 5.31% | W | 0.29% |
| q | 11.53% | U | 4.03% | T | 0.19% |
| ` | 10.50% | F | 2.60% | X | 0.08% |
| g | 10.33% | o | 2.12% | h | 0.03% |
| n | 9.26% | P | 1.73% | | |

We have at first placed the most frequently used consonants in the flexible keys in circular order from the most flexible key to least, then we again arranged the keys of next flexible from the least flexible key to most in reversed manner. Finally it is done again from most frequent key to least. This arrangement is used to protect key jamming. Though this arrangement is not 100% successful but it can reduce 30% of key jamming. As key 1 is the most frequently and flexibly used key we have used it for symbols as symbols are also frequently used. In key 1 we have arranged the symbols according to their frequency. The order is ( **v > † > w > x > | > y > , > t > s > r > ‰ ª > Š > ~ > „** ) The vowels are less frequently used hence we have placed them in key 9 which is the least flexible key. We have arranged the vowels according to their appearance in Bengali language. That is ( **A > B > C > D > E > G > H > I > J > F** ) We have used key 0 (Zero) for space as after typing every word we need to put a space. In this way we have designed our proposed keypad layout.

## V. PROPOSED KEYPAD LAYOUT

Our proposed keypad layout after arranging the alphabet is shown in the following figure, Figure 8.

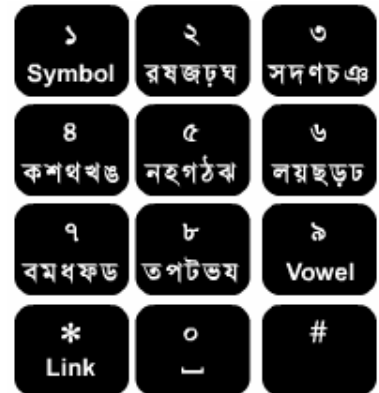

Figure 8: Our proposed frequency based keypad layout.

The key 1 represents the symbols ( **v > † > w > x > | > y > , > t > s > r > ‰ ª > Š > ~ > „** ), the key 9 represents the vowels ( **A > B > C > D > E > G > H > I > J > F** ). We have used key * for joining alphabets. If we want to type ( ' ) at first we have to type ( **b** ) and then press the link key * and then type ( **`** ).

## VI. ACHIEVEMENTS

The main achievement of this proposed method is to propose a standard cell phone keypad layout for not only the common people but also for this people who are nearly illiterate. In the proposed keypad layout the cell phone users will find the frequently used alphabet arranged in those keys, which are convenient to press, and does not provide very much pressure on the fingers. We have used key number 0 (Zero) for blank space, which will also be convenient one. As the user will find the most of the frequent alphabets in easily reachable keys, this arrangement will speed up the typing process. In our proposed keypad layout the most inconvenient key is 9 which we have used for vowels. It will lessen the finger works and will decrease the extra time and pressure required to move the position of the finger from the most flexible key to the least flexible one. Hence this frequency based keypad will be very much flexible to the cell phone users in comparison to other keypads.

## VII. CONCLUSIONS

For speeding up the typing process the alphabetic arrangement are required to be changed to greater extent. In our proposed keypad layout as we have identified the flexible keys and arranged the frequently used alphabets in those keys, it will be convenient and less pressure creating on the finger works. Although this process will require some learning time, the users will be benefited to great extent as soon as they become experienced to it.